%% file: Max-GES_Quantum-accepted-arXiv-submitted.tex
\documentclass[a4paper,twocolumn,11pt,accepted=2022-10-17]{quantumarticle}
\pdfoutput=1
\usepackage[utf8]{inputenc}
\usepackage[english]{babel}
\usepackage[T1]{fontenc}
\usepackage{amsmath}
\usepackage{hyperref}

\usepackage{tikz}
\usepackage{lipsum}
\usepackage{ltxgrid}
\usepackage{bm,bbm}
\usepackage{amsmath,amssymb}
\usepackage{graphicx,amsthm,epstopdf,amsmath}
\usepackage{subfigure,color}
\usepackage[normalem]{ulem}
\input{komendy_osid.tex}

\newtheorem{thm}{Theorem}

\newtheorem{fakt}[thm]{Fact}
\newtheorem{prop}[thm]{Proposition}

\newtheorem{obserwacja}[thm]{Observation}

\newcommand{\beu}{\begin{equation}}
\newcommand{\eeu}{\end{equation}}

\newcommand{\be}{\begin{eqnarray}}
\newcommand{\ee}{\end{eqnarray}}

\newcommand{\ba}{\begin{array}}
\newcommand{\ea}{\end{array}}

\newcommand{\cee}[1]{\mathbb{C}^{#1}}

\begin{document}

\title{Universal construction of genuinely entangled subspaces of any  size}
\author{Maciej Demianowicz}\orcid{0000-0001-7018-582X}
 \affiliation{{\it\small Institute of Physics and Applied Computer Science, Faculty of Applied Physics and Mathematics,
		Gda\'nsk University of Technology, Narutowicza 11/12, 80-233 Gda\'nsk, Poland}}
\email{maciej.demianowicz@pg.edu.pl}

\begin{abstract}
We put forward a simple  construction of genuinely entangled subspaces  -- subspaces supporting only genuinely multipartite entangled states -- of any permissible dimensionality for any number of parties and local dimensions.  The method uses nonorthogonal product bases, which are built from totally nonsingular matrices with a certain structure. We give an explicit basis for the constructed subspaces.
An immediate consequence of our result is the possibility of constructing in the general multiparty scenario genuinely multiparty entangled mixed states with ranks up to the maximal dimension of a genuinely entangled subspace.
\end{abstract}
\maketitle

\section{Introduction } 
	Genuinely multipartite entanglement (GME) \cite{gme-definicja} constitutes the strongest form of entanglement in many body quantum systems with a plethora of applications, e.g., teleportation and dense coding \cite{gme-teleport}, metrology \cite{toth-metro}, quantum key distribution \cite{epping}, or device independent cryptography \cite{gme-crypto}. For this reason, its characterization belongs naturally to the major study subjects of quantum information science. A particular trend within this branch, gaining recently considerable attention, focuses on the study of genuinely entangled subspaces (GESs) \cite{cubitt,upb-to-ges} (see also \cite{Parthasarathy}), i.e., subspaces composed only of GME states, which have been identified as objects of both theoretical and practical significance \cite{Agrawal,4x4,huber-grassl,nonlocal,DIsubspaces,condition}.
One of the most important research lines aims at providing their constructions. 
First systematic approach to the problem proposed in Ref. \cite{upb-to-ges} was based on unextendible product bases (UPBs) \cite{upb,big-upb} -- a natural notion for constructing entangled subspaces -- specifically, their nonorthogonal variant (nUPBs) \cite{pittenger}. Further developments made use of various tools and concepts: unextendible biproduct bases \cite{Agrawal}, certain characterization of bipartite completely entangled subspaces \cite{approach}, stabilizer formalism \cite{benchmarks,DIsubspaces2,npt-ges}, correspondence between quantum channels and subspaces of tensor product Hilbert spaces \cite{antipin}, or compositional (tensor product) approach \cite{antipin2}.
While some of these proposal \cite{approach,antipin} offer the possibility of constructing GESs of any sizes, their actual general utility  is only theoretical as they do not provide a recipe to achieve this task in any multipartite scenario and their applicability is in fact very limited and boils down to small systems. 
It has thus remained a recognized open problem to find a construction which is universal, i.e., 
gives a practical prescription for building GESs of any, including the maximal, dimensions in the most general multipartite scenario with any number of parties and arbitrary local dimensions. 
\newline \indent
It is the goal of the present paper to provide such a construction. With this aim we exploit totally nonsingular matrices to build nUPBs, which are biproduct unextendible and as such give GESs in their orthocomplements. The construction is simple and avoids the machinery of algebraic geometry commonly used in the study of entangled subspaces.
Finding explicit characterization of the subspaces requires only solving  homogeneous systems of linear equations, which allows us to give an explicit basis for the constructed subspaces.
Our result automatically ensures a construction of GME mixed states with varying ranks. 
Noteworthy, the proposed approach is  flexible as it allows for constructing different classes of GESs in a given multipartite setup, and as such provides a useful testbed for the theory of multipartite entanglement.

\section{Preliminaries } \label{preliminaries}
We consider $n$--partite Hilbert spaces 
$\calH_{d_0d_1\dots d_{n-1}}=\bigotimes_{i=0}^{n-1}\cee{d_i}$.
% if there are $n$ subsystems. 
Subsystems are denoted as $A_0A_1\dots A_{n-1}=: \bf{A}$ and
% or simply $A,B,C$ when there are three parties. 
 states as $\ket{\psi}_{X},\ket{\varphi}_{X},\cdots, \varrho_X, \sigma_X,\dots$, where the subscript corresponds to parties holding them. Column vectors are written as $(x,y,\dots)$ in the main text, i.e., we omit the transposition; we denote $\ket{ab}:=\ket{a}\otimes\ket{b}$.

We call an $n$--partite pure state $\ket{\psi}_{\bf{A}}$  {\it fully product} if it can be written as $\ket{\psi}_{\bf{A}}=\ket{\varphi}_{A_0}\otimes \ket{\phi}_{A_1}\otimes\cdots\otimes \ket{\xi}_{A_{n-1}}$. Otherwise, it is said to be {\it entangled}. An entangled multipartite state $\ket{\psi}_{\bf{A}}$, which cannot be written in a {\it biproduct} form, i.e., $\ket{\psi}_{\bf{A}}\ne\ket{\varphi}_{S}\otimes \ket{\phi}_{\bar{S}}$, for any bipartite cut (bipartition) $S | \bar{S}$, $S\cup \bar{S}=\bf{A}$, is said to be {\it genuinely multiparty entangled (GME)}. A mixed state $\varrho_{\bf{A}}$ is called GME if it is not {\it biseparable}, i.e., $\varrho_{\bf{A}} \ne \sum_{S|\bar{S}} p_{S|\bar{S}} \sum_i q^{(i)}_{S|\bar{S}} \rho^{(i)}_{S} \otimes \sigma^{(i)}_{\bar{S}}$, where the first sum goes over all bipartitions.

 The notions defined for single pure states transfer naturally to subspaces. A subspace $\calS \subset \calH$ is then called a {\it completely entangled subspace (CES)} iff all states belonging to it are entangled \cite{Parthasarathy,Bhat,WalgateScott}.   
 Further, if all states from $\calS  \subset \calH$ are GME, then $\calS$ is called a {\it genuinely entangled subspace (GES)} \cite{upb-to-ges}. The maximal dimension of a GES of %\textcolor{red}{$\calH_{d^n}$ equals  $ (d^{n-1}-1)(d-1)$} 
  $\calH_{d_0d_1\dots d_{n-1}}$ equals
 $D-D/d_{\mathrm{min}}-d_{\mathrm{min}}+1$ with $D=\Pi_{i=0}^{n-1}d_i$ and $d_{\mathrm{min}}=(d_0,d_1,\dots,d_{n-1})$; %
for equal local dimension, $d_i=d$, this simply gives
 $ (d^{n-1}-1)(d-1)$.

 A common approach to the construction of entangled subspaces is to use unextendible product bases (UPBs). A UPB is a set of fully product vectors such that there is no other fully product vector orthogonal to all of them \cite{upb,big-upb,pittenger,alon-lovasz, johnston-qubit}. Depending on whether the vectors of a UPB are mutually orthogonal or not, we speak of orthogonal (oUPBs) or nonorthogonal UPBs (nUPBs), respectively.
Clearly, in both cases, by definition, the orthocomplement of a subspace spanned by  a UPB is a CES. An elementary lower bound on the cardinality of a UPB is $\sum_{i=0}^{n-1}d_i-n+1.$ A multipartite UPB with the additional property that it is a UPB for {\it any} bipartition of the parties leads to a GES and there must be at least $d_{\mathrm{min}}+D/d_{\mathrm{min}}-1$ in such a set. This observation allowed us to build (large but not maximal) GESs  as orthocomplements  of subspaces spanned by nUPBs \cite{upb-to-ges}. The core of the construction was the following beautiful result about local spanning \cite{upb,big-upb}, applied to all bipartitions.
\begin{fakt}\label{glowny-fakt} 
Given is a set of product vectors from $\mathbb{C}^m \otimes \mathbb{C}^n$: $B=\{\ket{\varphi_i}\otimes \ket{\psi_i}\}_{i=1}^k$  with $k \ge m+n-1$. If any $m$ vectors $\ket{\varphi_i}$ span $\mathbb{C}^m$ and any $n$ vectors $\ket{\psi_i}$ span $\mathbb{C}^n$, then there does not exist a product vector  orthogonal to all the vectors from $B$.
	\end{fakt}

 \noindent Vectors $\ket{\varphi_i}$ and $\ket{\psi_i}$ with the property as in the fact above are said to possess the {\it spanning property}.  One can easily see that while Fact \ref{glowny-fakt} is in general only a sufficient condition for unextendibility, it is also necessary for the minimal cardinality, $k=m+n-1$.

A weaker version of Fact \ref{glowny-fakt}, which is both necessary and sufficient for any cardinality, states:  {\it let $P$ be a partition of  $B$ into two disjoint sets: $B = B_1 \cup B_2$; $B$ is extendible iff there exist a partition $P$ such that the local vectors on the first site from $B_1$ do not span $\mathbb{C}^m$ and the local vectors on the second site from $B_2$ do not span $\mathbb{C}^n$.}

Our another important tool are totally nonsingular (TNS) matrices, that is matrices whose all minors are nonvanishing.  By definition, such matrices do not have zero entries. A class of TNS matrices is comprised of totally positive (TP) ones, i.e., those whose all minors are strictly positive. A simple, yet important, fact holds.
\begin{fakt}
	\label{tns} Given a TNS matrix $[A]_{ij}=a_{ij}$, matrices constructed from $A$ by multiplying its rows or columns by nonzero constants,  $[\tilde{A}]_{ij}=h_i a_{ij}$, $h_i \ne 0$, and  $[\bar{A}]_{ij}=g_j a_{ij}$ , $g_j\ne 0$, are also TNS.
	\end{fakt}
%
%\sout{Clearly, a more general statement follows that all (rectangular) submatrices of such matrices are full rank.} 

\section{Construction}

\begin{figure}[h!]
	\includegraphics[width=8cm,height=3.2cm]{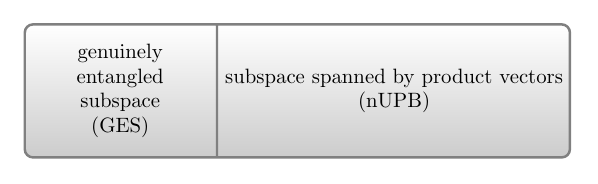}
	\caption{Genuinely entangled subspaces (GESs) are constructed as the orthocomplements of the span of  fully product vectors. %\textcolor{red}{$\ket{\Psi^{(i)}}_{\bf{A}}$ [cf. \eqref{wektory}] or, more generally, $\ket{\tilde{\Psi}^{(i)}}_{\bf{A}}$ [cf. (\ref{wektory-ogolniej})]}. 
		These vectors, which are built from %\textcolor{red}{elements of discrete Fourier transform (DFT) matrices of prime order} 
		totally nonsingular matrices,  form nonorthogonal unextendible product bases (nUPBs).     
		Their additional property leading to GESs is that the unextendibility property holds for any bipartition, 
		 which follows from Fact \ref{glowny-fakt}. Explicit characterization of a subspace involves solving a homogeneous system of linear equations.
		The construction is valid for any multipartite Hilbert space and allows for constructing GESs with arbitrary dimensions.
		}\label{decomposed}
\end{figure}

We now present our construction, which is based on nUPBs and Fact \ref{glowny-fakt} (see Fig. \ref{decomposed}).
In general, the  challenging part in such approach is ensuring the spanning property to hold for any bipartition. Here, we achieve this by constructing bases from totally nonsingular matrices for which this property will be satisfied automatically. The procedure is inspired by one from Ref. \cite{big-upb}, where the properties of a class of such matrices were used  to build minimal oUPBs in certain bipartite systems. %\textcolor{red}{without the need to check it}. 

Before we proceed, let us briefly give reason why we assume from the very beginning that the basis vectors are nonorthogonal. The use of nUPBs instead of oUPBs is necessary to have a construction working universally in any multiparty scenario. First, if we dropped this assumption, we would not be able to take into account the case of qubit subsystems since, as it is well known, oUPBs do not exist in $2 \otimes m$ systems \cite{upb}. Moreover,  it has been recently shown that in many setups there exists a nontrivial bound on the cardinalities of oUPBs unextendible with biproduct vectors \cite{no-gupb}. 

The method is summarized in the following proposition.
\begin{prop}\label{main-result}
	Given is a set of $K\ge D/d_{\mathrm{min}}+d_{\mathrm{min}}-1$ fully product (unnormalized) states from $\calH_{d_0d_1\dots d_{n-1}}$: 
	\begin{equation}
		\label{wektory-general}
		\ket{\Psi^{(i)}}_{\bf{A}}=\bigotimes_{m=0}^{n-1} \ket{\xi_m^{(i)}}_{A_m}, \quad i=0,1,\dots,K-1.
	\end{equation}
	%
%$K \ge D/d_{\mathrm{min}}+d_{\mathrm{min}}-1$.
	%and construct a matrix with rows being the product vectors (\ref{wektory-general}):
	Let
	\begin{equation}\label{matrix-M}
		M=\sum_{i=0}^{K-1} \left(\ket{\Psi^{(i)}}_{\bf{A}}\bra{i}\right)^T.
	\end{equation}
If  $M$ is TNS, then $\left(\spann\:\{\ket{\Psi^{(i)}}_{\mathrm{\textbf{A}}}\}_{i=0}^{K-1}\right)^{\perp}=\mathrm{null}(M)$ is a GES of dimension $D-K$. 
%In other words, the (right) null space of $M$, $\mathrm{null}(M)$,  is a GES.
	\end{prop}
\begin{proof} If $M$ is TNS, then it is full rank meaning that $\ket{\Psi^{(i)}}_{\textbf{A}}$'s are linearly independent and  $\dim (\mathrm{null}(M))=D-K$.
 Consider now a bipartition $S|\bar{S}$ of $\bf{A}$ and write $\ket{\Psi^{(i)}}_{\bf{A}}=\ket{\varphi^{(i)}}_S\otimes\ket{\phi^{(i)}}_{\bar{S}}$ with $\ket{\varphi^{(i)}}_S=(a_0^{(i)},a_1^{(i)},\dots)_S$ and $\ket{\phi^{(i)}}_{\bar{S}}=(b_0^{(i)},b_1^{(i)},\dots)_{\bar{S}}$. Let 
\begin{align}\label{lokal-N}
& N_S^{(p)} =\sum_{i=0}^{K-1} b_p^{(i)}\left(\ket{\varphi^{(i)}}_S\bra{i}\right)^T,\\
& N^{(p)}_{\bar{S}} = \sum_{i=0}^{K-1} a_p^{(i)} \left(\ket{\phi^{(i)}}_{\bar{S}}\bra{i}\right)^T
\end{align}
for some $p$; these matrices are visibly submatrices of $M$. 
%Now,if $M$ is a submatrix of a TNS matrix, then
Further, all submatrices of both $N_S^{(p)}$ and $N^{(p)}_{\bar{S}}$ are TNS. By Fact \ref{tns} ($a_p^{(i)}\ne 0$, $b_p^{(i)}\ne 0$ for all $p$'s and $i$'s) this is also true for the matrices $\sum_{i=0}^{K-1} \left(\ket{\varphi^{(i)}}_S\bra{i}\right)^T$ and $\sum_{i=0}^{K-1} \left(\ket{\phi^{(i)}}_S\bra{i}\right)^T$ and
it follows that $\{\ket{\varphi^{(i)}}_S\}$ and $\{\ket{\phi^{(i)}}_{\bar{S}}\}$ possess the spanning property. This is true for any bipartition $S|\bar{S}$ implying by Fact \ref{glowny-fakt} that vectors \eqref{wektory-general} form a set, which is unextendible for any bipartition, i.e., there does not exist a biproduct vector orthogonal to $\spann\:\{\ket{\Psi^{(i)}}_{\textbf{A}}\}_{i=0}^{K-1}$. 
The claim follows.
\end{proof}

	To get an explicit characterization of a subspace given in this manner we simply look for the null space of $M$. This amounts to solving  a homogeneous system of $K$ linear equations on $D$ variables (thus with $D-K$ free parameters, which can be chosen upon convenience).
	The obtained basis will  be nonorthogonal in general and the Gram--Schmidt procedure can be further utilized to orthogonalize it.

We now apply Proposition \ref{main-result} to construct explicitly GESs of arbitrary dimensions for any $\calH_{d_0d_1\dots d_{n-1}}$. With this purpose we utilize the ubiquitous Vandermonde matrices, whose usefulness in analyses of entangled subspaces was already recognized in the past \cite{cubitt,Parthasarathy,Bhat,skowronek}.

%\subsection{Construction I: Vandermonde matrices}
A Vandermonde matrix is a matrix of the form:
\begin{equation}\label{vandermonde-matrix}
	V_{m,n}(\textbf{x})=\left(\begin{array}{ccccc}
	1	& x_0 & x_0^2  & \dots & x_0^{n-1} \\
	1	& x_1 & x_1^2  & \dots & x_1^{n-1}   \\
	\vdots	& \vdots  & \vdots  & \ddots  & \vdots \\
		1	& x_{m-1} & x_{m-1}^2  & \dots & x_{m-1}^{n-1} 
	\end{array} \right);
\end{equation}
numbers $x_i$ are called the nodes of $V_{m,n}$. Matrices with elements $x_i^{a_{j}}$ in the $i$--th row, where $a$ is a sequence of nonnegative integers, are called generalized Vandermonde matrices
%, $V_{m,n}^{G_{a}}$; any $V_{M,N}^{G_a}$ 
and any such matrix is a submatrix of some $V_{m,n}$.
% A submatrix of $V_{m,n}$ built from rows and columns indexed by, respectively, $\textbf{r}$ and $\textbf{c}$, is called a generalized Vandermonde matrix, $V_{m,n}^{G_{\textbf{r},\textbf{c}}}$. 
A particular example of $V_{m,n}$ is the DFT matrix of order $p$, which is given by (we omit the constant factor):
\begin{align}
&DFT_p=[\omega^{mn}]_{m,n=0}^{p-1}\\
&\hspace{+0cm}=\left( \begin{array}{ccccc}
	1	&1  & 1   & \dots  & 1 \\
	1	& \omega  & \omega^2   & \dots  & \omega^{p-1} \\
	1	& \omega^2 & \omega^4 & \dots & \omega^{2(p-1)} \\
	1	& \omega^3 & \omega^6 & \dots & \omega^{3(p-1)} \\
	\vdots	& \vdots & \vdots  & \ddots & \vdots \\
	1	& \omega^{p-1} & \omega^{2(p-1)} & \dots  & \omega^{(p-1)(p-1)}
\end{array} \right)\nonumber ,
\end{align}
where $\omega$ is a primitive $p$th root of unity, $\omega=\eksp^{2\pi\uroj/p}$, that is $x_j=\omega^{j}$.
\newline\indent
Crucially for our construction, it holds that
\begin{itemize} 
	\item[(i)] $V_{m,n}$ with positive nodes $0<x_0<x_1<\dots<x_{m-1}$ is TP, 
	\item[(ii)] $DFT_p$ with prime $p$ is TNS (this result is known as the Chebotarev theorem on roots of unity \cite{chebotarev}; see, e.g., \cite{tao} for a  recent proof).
\end{itemize}
\indent\indent Further, notice that rows of $V_{l,D}$ have a tensor product structure when treated as vectors from $\calH_{d_0d_1\dots d_{n-1}}$:
\begin{align}\label{vandermonde-vector}
&\ket{v_{d_0d_1\dots d_{n-1}}(x)}_{\textbf{A}}:=\left(1,x,x^2,\dots,x^{D-1}\right)_{\textbf{A}} \non
&\hspace{+0.5cm}=\bigotimes_{m=0}^{n-1} \left( \sum_{s_m=0}^{d_m-1} x ^ {q_{m,s_m}}\ket{s_m} \right)_{A_m}\\
&\hspace{+0.5cm}=  \sum_{s_0=0}^{d_0-1}\cdots \sum_{s_{n-1}=0}^{d_{n-1}-1} \displaystyle x ^{\sum_{m=0}^{n-1}q_{m,s_m}}\ket{s_0s_1\cdots s_{n-1}}_{\textbf{A}},\nonumber
\end{align}
where
\begin{equation}\label{q}
q_{m,s_m}=s_m \prod_{k=m+1}^{n-1}d_k,
\end{equation}
$m=0,1,\dots,n-1$, $s_m=0,1,\dots,d_m-1$.
The numbers \eqref{q} follow from the conversion of $s_0s_1\dots s_{n-1}$ to the decimal representation.
\newline\indent
We now have all the elements needed to prove the following.
\begin{obserwacja}\label{vandermonde-case}
Given is a set of vectors from $\calH_{d_0d_1\dots d_{n-1}}$:
	\begin{align}
	&\hspace{0cm}	\ket{\Psi_{\textbf{A}}^{(i)}}= \ket{v_{d_0d_1\dots d_{n-1}}(x_i)}_{\textbf{A}}, \\
	&\hspace{+0.7cm}	i=0,1,\dots, K-1, \;\; K \ge  D/d_{\mathrm{min}}+d_{\mathrm{min}}-1. \nonumber
	\end{align}
	If
	\begin{itemize}
		\item[(i)]  $0<x_0<x_1< \dots < x_{K-1}$
		\end{itemize}
	or
	\begin{itemize}
		\item[(ii)] $x_i=\omega^i$, where $\omega=\eksp^{2\pi\uroj/p}$ with prime $p>D$,
	\end{itemize}
then $\left(\spann\:\{\ket{\Psi^{(i)}}_{\textbf{A}}\}_{i=0}^{K-1}\right)^{\perp}$ is a GES of dimension $D-K$.
\end{obserwacja}
\begin{proof}  This follows straightforwardly from Proposition \ref{main-result}. The matrix $M$  [cf. Eq. \eqref{matrix-M}] is (i)  a TP Vandermonde matrix, $V_{K,D}(\textbf{x})$, or (ii) a submatrix of a DFT matrix of prime order $p$, and as such it is TNS.
\end{proof}
The problem of finding an explicit characterization of the GESs from Observation \ref{vandermonde-case} boils down to determining the null space of $V_{K,D}(\textbf{x})$ corresponding to (i) and (ii).
 Importantly, this can be easily achieved with the  use of the inverse of a Vandermonde matrix \cite{inverse-vandermonde} and we find that $\mathrm{null}(V_{K,D}(\textbf{x}))$ is spanned by (see Appendix \ref{kernel-V})
 \begin{align}\label{baza-ges}
 & \ket{g^{(\alpha)}}\\
&	\hspace{-0.1cm}=\sum_{i,j=0}^{K-1} \frac{(-)^{K-i}\sigma_{K-i-1}(\textbf{x}\setminus x_j)}{\prod_{l=0,l\ne j}^{K-1}(x_j-x_l)}x_j^{K+\alpha}\ket{i}+\ket{K+\alpha}, \non
& \hspace{4.05cm}	\alpha=0,1,\dots,D-K-1, \nonumber
 \end{align}
 where $\textbf{x}=(x_0,\dots,x_{K-1})$, $\ket{j}=\ket{s_0s_1\dots s_{n-1}}$, $j=\sum_{m=0}^{n-1}s_m \prod_{k=m+1}^{n-1}d_k$, and
 \begin{equation}\label{symmetric-polynomial}
 	\sigma_k(a_0,\dots,a_{N-1})=\sum_{0 \le j_0  < \dots < j_{k-1} \le N-1} \hspace{-0.1cm} a_{j_0}  \cdots  a_{j_{k-1}} 
 \end{equation}
 is the $k$th symmetric polynomial ($\sigma_0\equiv 1$).
We can thus state the following
\begin{obserwacja}
Subspace $\spann \{\ket{g^{(\alpha)}}\}_{\alpha=0}^{D-K-1}$ [cf. Eq. \eqref{baza-ges}] with (i) $0<x_0<\dots<x_{K-1}$ or (ii) $x_i=\omega^i$, where  $\omega=\eksp^{2\pi\uroj/p}$ with prime $p>D$, is a GES od dimension $D-K$.
\end{obserwacja}

We observe that we can use more general nUPBs to build GESs:
	%We can use Fact \ref{tns} further and consider nUPBs  more general  than those given by (\ref{wektory}). More precisely, one can use the following vectors in the construction:
	%
	\begin{equation}\label{wektory-ogolnie}
	\ket{\tilde{\Psi}^{(i)}}_{\bf{A}}= \bigotimes_{m=0}^{n-1} \left( \sum_{s_m=0}^{d_m-1} h_{m,s_m}x_i ^ {\tilde{q}_{m,s_m}^{(i)}}\ket{s_m} \right)_{A_m},
	\end{equation}
	where $\tilde{q}_{m,s_m}^{(i)}$'s are chosen in such a way that the matrix $M$ is a generalized Vandermonde matrix, and
	 $h_{m,s_m}> 0$ (irrelevance of local unitaries allows us to assume this) with  $h_{m,s_0}=1$ (unimportance of a global factor). 
Let us finally note that the requirement on $p$ to be prime in the construction based on $DFT_p$ is not a necessary condition. There do exist GESs constructed in the same manner for nonprime $p$'s (for example $p=9$ in the three-qubit case), however, verifying when this is the case in general might require the application of the weaker version of  Fact \ref{glowny-fakt} discussed in Section \ref{preliminaries}, which would  be very hard in an arbitrary  multipartite setting.

	We thus see that the  approach aside from its universality offers  a lot of generality, which could potentially be further exploited in constructions of genuinely entangled subspaces and states with certain properties.

\subsection{Examples}\label{examples}
Here we give  examples of the construction from Observation \ref{vandermonde-case} in the case of three qubits. %For simplicity we focus on three qubits, in which case the nUPBs have five vectors.
\newline\indent
The following nUPBs give rise to maximal GESs:
\begin{equation}\label{3kubity-vandermonde}
%&&\ket{\tilde{\Psi}^{(i)}}_{\textbf{A}}= (1,(i+1)^8)_{A_0}\otimes (1,(i+1)^4)_{A_1} \otimes (1,(i+1)^2)_{A_2} \otimes (1,i+1)_{A_3}, \non
\ket{\bar{\Psi}^{(i)}}_{\textbf{A}}=\bigotimes_{m=0}^2 \left(1, (i+1)^{2^{2-m}}\right)_{A_m} ,
\end{equation}
$i=0,1,\dots,4$,	where we have set the nodes of the Vandermonde matrix to be $x_i=i+1$, and
\begin{align}\label{3kubity-dft}
&\hspace{-0.3cm}\ket{\Psi^{(0)}}_{\textbf{A}}= (1,1)_{A_0}\otimes (1,1)_{A_1} \otimes (1,1)_{A_2}, \nonumber \\
&\hspace{-0.3cm}\ket{\Psi^{(1)}}_{\textbf{A}}=(1,\omega^{4})_{A_0}\otimes (1,\omega^{2})_{A_1} \otimes (1,\omega^{1})_{A_2},  \nonumber\\
&\hspace{-0.3cm}\ket{\Psi^{(2)}}_{\textbf{A}}=(1,\omega^{8})_{A_0}\otimes (1,\omega^{4})_{A_1} \otimes (1,\omega^{2})_{A_2}, \\
&\hspace{-0.3cm}\ket{\Psi^{(3)}}_{\textbf{A}}=(1,\omega^{12})_{A_0}\otimes (1,\omega^{6})_{A_1} \otimes (1,\omega^{3})_{A_2},  \nonumber\\
&\hspace{-0.3cm}\ket{\Psi^{(4)}}_{\textbf{A}}=(1,\omega^{16})_{A_0}\otimes (1,\omega^{8})_{A_1} \otimes (1,\omega^{4})_{A_2}, \nonumber
\end{align}
where we can choose $\omega =\eksp ^{2\pi \uroj/11}$, i.e., $p=11$.
	\newline\indent
 Basis vectors for both GESs are given in Appendix \ref{explicit-3q-GES}.

\subsection{GME mixed states}\label{gme-mixed}

While it is in principle easy to verify if a pure state is GME (by looking at the reduced density matrices for different bipartitions), this problem is far from trivial for mixed states (see, e.g., \cite{gme-criterion,taming,murta,gme-uncertainty}). It is thus important to be able  to construct states whose entanglement properties we have the knowledge of. Entangled subspaces allow for this task -- any state supported on such a subspace is necessarily entangled with the type of entanglement corresponding to the type of the entangled subspace.
This follows from the fact that different ensembles for a mixed state are related through isometries \cite{ansamble}. With GESs of any dimensions in the general multipartite setup in hand we can construct GME states with ranks up to the maximal one within the approach. 
Particularly significant states from this class are the normalized projections onto subspaces.
It is of interest to investigate  in more detail the entanglement of such states, in particular, how it behaves under mixing with the white noise (cf. \cite{ent-ges}) as this might give a further characterization of the GESs considered here and their usefulness in information processing tasks. This  is, however, beyond the scope of the present paper. 

An important class of entangled mixed states is comprised of bound entangled (BE) states with positive partial transposition (PPT). It is well known that oUPBs are a natural and direct 	tool for their construction --  a normalized projection onto the orthocomplement of a subspace spanned by an oUPB is a PPT BE state  \cite{upb}. On the other hand, as already noted in \cite{big-upb},  there is no reason to expect that nUPBs could also serve this purpose in general  (albeit see Refs. \cite{skowronek,leinaas,chen-djokovic} for a connection between bipartite nUPBs and low--rank low dimensional PPT BE states).
This obviously does not exclude the possibility that in some settings this construction may work, or, more generally, a given subspace supports PPT BE states, but they need to be constructed in a more sophisticated manner. We have checked numerically that is not true for the nUPBs from Proposition \ref{vandermonde-case} with some chosen nodes for low dimensional small systems, but
we have not been able to verify whether this holds in general for the given subspaces.
 A somewhat complementary question is whether our approach allows for a construction of fully nonpositive partial transpose (NPT) GESs, that is GESs that support only states being NPT across all bipartitions \cite{Agrawal,antipin2}, and in case of a positive answer which dimensions of GESs could be achieved. This is especially intriguing in view of the recent construction of large fully NPT stabilizer GESs  \cite{npt-ges}.  We leave both problems for future study as it seems that without a fairly simple explicit orthogonal bases for the subspaces both problems may turn out analytically formidable.

\section{Conclusions and outlook}

We have shown how to construct genuinely entangled subspaces (GESs) of any permissible size in any multipartite setting.  The construction is based on nonorthogonal unextendible product bases (nUPBs), which are built upon  totally nonsingular matrices.  We provide an explicit basis for such constructed subspaces.
Through a standard argument our result allows for a construction of genuinely multiparty entangled (GME) mixed states of arbitrary ranks no greater than the maximal possible dimension of a GES.  These states might be a useful tool to benchmark multiparty entanglement criteria.

 The use of nUPBs was dictated by the goal of providing a construction working for any multipartite Hilbert space and any dimensions of GESs, which is known not to be achievable with orthogonal UPBs (oUPBs) \cite{upb,no-gupb}.
In fact, despite some progress reported very recently in  \cite{tiles-gupb}, it still remains a big open problem whether oUPBs could be used at all in the construction of GESs. In view of our approach, a question arises of what class of matrices could be  used potentially in an attempt to solve the problem. Its positive resolution would have important consequences as this is related to the construction of  bound entangled states and  sets of product states indistinguishable for any bipartition.   We believe our contribution may stimulate research within this direction.
%\textcolor{red}{On a less fundamental level, there remains open the problem of constructing GESs of arbitrary sizes with simple explicit orthogonal bases.}
On another note, our work leaves open the problem of finding a universal construction of GESs with (simple) explicit orthogonal bases. Their availability would facilitate greatly   analyses of the arising subspaces and states.

\onecolumn
%\newpage
\appendix
\section{The null space of a Vandermonde matrix}\label{kernel-V}
Here we show how to find $\mathrm{null}\left(V_{m,n}\right)$ with $V_{m,n}$ given by Eq. \eqref{vandermonde-matrix}; we assume $m<n$ so that the null space is nontrivial. We need to 
find $n-m$ linearly independent solutions
of 
\begin{equation}\label{kernelowe}
V_{m,n} \ket{a^{(s)}}=0, \quad s=0,1,\dots,n-m-1.
\end{equation}
Letting $\ket{a^{(s)}}=\sum_{i=0}^{n-1}a^{(s)}_i \ket{i}$, % for all $\ket{a}$'s. 
Eq. \eqref{kernelowe} gives a homogeneous system of linear equations for each $s$:
\begin{equation}
	\sum_{j=0}^{n-1} x_i^j a_j^{(s)}=0,\quad i=0,1,\dots,m-1.
\end{equation}
We choose 
\begin{equation}
	a_i^{(s)}=\delta_{i-m,s},\quad i=m,m+1,\dots,n-1
\end{equation}
 to build a basis for  $\mathrm{null}\left(V_{m,n}\right)$ and with such a choice we obtain for each $s$:
\begin{equation} \label{free-parametry}
\sum_{j=0}^{m-1} x_i^j a_j^{(s)}=-x_i^{m+s},
\quad i=0,1,\dots,m-1.
\end{equation}
Equivalently, we have
\begin{equation}
	V_{m,m}\ket{a^{(s,m)}} = \ket{\mathcal{X}^{(s)}},
\end{equation}
where $\ket{a^{(s,m)}}=\sum_{i=0}^{m-1}a_i^{(s)} \ket{i}$ and $\ket{\mathcal{X}^{(s)}}=\sum_{i=0}^{m-1} (-x_i^{m+s})\ket{i}$. Its solution is given as
\begin{equation}\label{odwrotne}
	\ket{a^{(s,m)}}=	V_{m,m}^{-1}\ket{\mathcal{X}^{(s)}}.
\end{equation}
\newline\indent
The inverse of $V_{m,m}(\textbf{x})$ is found to be \cite{inverse-vandermonde}
\begin{equation}\displaystyle
	V_{m,m}^{-1}=\sum_{i,j=0}^{m-1}\frac{(-)^{m-i-1}\sigma_{m-i-1}(\textbf{x} \setminus x_j)}{\prod_{k=0,k\ne j}^{m-1} (x_j-x_k)} \ket{i}\bra{j},
\end{equation}
where $\textbf{x}=(x_0,x_1,\dots,x_{m-1})$ with $x_i\ne x_j$ for $i\ne j$, and $\sigma_k$ is the $k$th symmetric polynomial [cf. Eq. \eqref{symmetric-polynomial}].
Inserting this into Eq. \eqref{odwrotne} and taking into account our choice of free parameters in Eq. \eqref{free-parametry} we obtain:
\begin{align}
	\ket{a^{(s)} }= \sum_{i,j=0}^{m-1}\frac{(-)^{m-i}\sigma_{m-i-1}(\textbf{x} \setminus x_j)}{\prod_{k=0,k\ne j}^{m-1} (x_j-x_k)} x_j^{m+s} \ket{i}+\ket{m+s}, \quad \quad
	 s=0,1,\dots,n-m-1. 
\end{align}
Orthogonalization of the basis can be performed with standard tools.

\section{Examples of maximal GESs: three-qubit case} \label{explicit-3q-GES}
An  (unnormalized) orthogonal basis for the GES given by vectors \eqref{3kubity-vandermonde}:
%
%\begin{widetext}
	\beq%\label{}
%	\hspace{-0.3cm}
	\ket{\bar{\Phi}^{(1)}}_{\textbf{A}}=
	\begin{pmatrix}
	-120 \\ 274 \\ -225\\85 \\ -15\\ 1 \\0\\ 0
	\end{pmatrix},
	\ket{\bar{\Phi}^{(2)}}_{\textbf{A}}= 
	\begin{pmatrix}
	-4\;597\;800 \\ 4\;596\;230 \\ 4\;855\;541 \\ -7\;809\;625 \\ 3\;605\;915 \\ -699\;445 \\ 49\;184 \\0
	\end{pmatrix},
	\ket{\bar{\Phi}^{(3)}}_{\textbf{A}}=
	\begin{pmatrix}
	-211\;257\;321\;000 \\ 51\;088\;047\;350\\ 246\;285\;410\;585\\ 122\;729\;176\;307 \\
	-362\;858\;132\;500\\ 190\;361\;914\;130\\ -39\;202\;174\;765\\ 2\;853\;079\;893
	\end{pmatrix} .
	\eeq
%\end{widetext}
%
	An orthonormal  basis for the GES given by vectors \eqref{3kubity-dft}: 
%
%\begin{widetext}
	\begin{align}
	%\label{}
	%\hspace{-0.3cm}
	\ket{\Phi^{(1)}}_{\textbf{A}} \hspace{-0.1cm}=\hspace{-0.1cm}
	\begin{pmatrix}
 -0.0856+0.0550 \uroj \\ -0.0509-0.3539 \uroj \\ 0.5772+0.1695 \uroj \\ -0.3939+0.4546 \uroj \\ -0.1485-0.3252 \uroj \\ 0.1018\\ 0 \\ 0\\
	\end{pmatrix}\hspace{-0.1cm},\:
	\ket{\Phi^{(2)}}_{\textbf{A}}\hspace{-0.1cm}= \hspace{-0.1cm}
	\begin{pmatrix}
-0.1474-0.0947 \uroj \\ 0.3473-0.2232 \uroj \\ 0.0461+0.3205 \uroj \\ 0.1555+0.0457 \uroj \\ -0.3812+0.4399 \uroj \\ -0.2229-0.4881 \uroj \\  0.2026 \\ 0
	\end{pmatrix}\hspace{-0.1cm},\:
	\ket{\Phi^{(3)}}_{\textbf{A}}\hspace{-0.1cm}=\hspace{-0.1cm}
	\begin{pmatrix}
	0.0352-0.2448 \uroj \\ 0.3185+0.2047 \uroj \\ -0.0497+0.0319 \uroj \\ 0.0446+0.3099 \uroj \\ -0.1338-0.0393 \uroj \\ -0.2715+0.3133 \uroj \\ -0.2629-0.5757 \uroj \\ 0.3197
	\end{pmatrix} \hspace{-0.1cm}.
	\end{align}
%\end{widetext}
%

\end{document}

%% file: komendy_osid.tex
\newcommand{\beq}[0]{\begin{equation}}
\newcommand{\eeq}[0]{\end{equation}}
\newcommand{\bw}[0]{\begin{widetext}}
\newcommand{\ew}[0]{\end{widetext}}
\newcommand{\bc}[0]{\begin{center}}
\newcommand{\ec}[0]{\end{center}}
\newcommand{\bwn}[0]{\begin{widetext}\begin{eqnarray}}
\newcommand{\ewn}[0]{\end{eqnarray}\end{widetext}}
\newcommand{\beqn}[0]{\begin{eqnarray}}
\newcommand{\eeqn}[0]{\end{eqnarray}}

\newcommand{\uroj}[0]{\mathrm{i}}
\newcommand{\eksp}[0]{\mathrm{e}}

\newcommand{\ket}[1]{|#1\rangle}
\newcommand{\bra}[1]{\langle #1 |}

\newcommand{\non}[0]{\nonumber\\}

\newcommand{\spann}[0]{\mathrm{span}}

\def\calH{{\cal H}}

\def\calS{{\cal S}}